\documentclass[pra, 12 pt]{revtex4}
\usepackage[english]{babel}
\usepackage{amsmath}
\usepackage{epsfig}

\begin{document}



\title{ON THE NECESSARY CONDITIONS FOR THE VALIDITY OF THE HOHENBERG-KOHN THEOREM}

\author{V.B. Bobrov $^{1}$, S. A. Trigger $^{1,\,2}$, G.J.F. van Heijst $^2$}
\address {$^1$ Joint Institute for High Temperatures, Russian Academy of
Sciences, IIzhorskaya St., 13, Bd. 2, 125412 Moscow, Russia; e-mail: satron@mail.ru\\
$^2$ Eindhoven  University of Technology, P.O. Box 513, MB 5600
Eindhoven, The Netherlands}
\date{4 July 2010}

\begin{abstract}
It is shown that the Hohenberg-Kohn lemma and theorem are direct consequences of the statement that the ground state energy (or free energy)
of a system of interacting particles in an external field is a unique functional of the potential of this field.
This means that, if the Hohenberg-Kohn theorem is valid, the nonuniform density in the equilibrium system and the external
field potential are biunique functionals. In this case, the nonuniform density is intimately related to the inverse response function.
On this basis, a regular procedure can be constructed for determining the density functional for the free energy or ground state energy.\\

PACS number(s): 71.15.Mb, 05.30.Ch, 71.10.Ca, 52.25.Kn \\

\end{abstract}

\maketitle In 1964, Hohenberg and Kohn  [1] have proved a theorem
according to which the ground state energy of inhomogeneous
electron gas in a static external field is a unique functional of
the nonuniform electron density. The Hohenberg-Kohn theorem is the
basis of the density functional theory widely used in various
fields of physics and chemistry [2-8]. The Hohenberg-Kohn theorem
proof is based on the following lemma [1,9]: the nonuniform
density $n(\bf r)$ in the ground state of a bounded system of
interacting electrons in a certain static external field
characterized by the potential  $\varphi^{ext}({\bf r)}$ uniquely
defines this potential. In this case [1,9]:

(a) The term "uniquely" means "with an accuracy up to an additive
constant of no interest";

(b) In the case of a degenerate ground state, the lemma relates to
the density  $n(\bf r)$  of any ground state. The requirement for
the ground state non-degeneracy is easily eliminated [10];

(c) The lemma is mathematically rigorous.

Let us pay attention to statement (c). The lemma proof is based on
the strict inequality [1,9]
\begin{eqnarray}
E_1<E_2 +\int [\varphi^{ext}_1 ({\bf r}) - \varphi^{ext}_2{\bf r}] n({\bf r}) d{\bf r} ,
\label{1}
\end{eqnarray}
Here $n({\bf r)}$   is the density of the non-degenerate ground state of the system of   electrons in the potential
$\varphi^{ext}_1 ({\bf r})$ corresponding to a state with wave function  $\Psi_1$ and energy $E_1$,
\begin{eqnarray}
E_1= \langle \Psi_1 | H_1 | \Psi_1 \rangle  =\int \varphi^{ext}_1 ({\bf r}) n({\bf r}) d{\bf r}+
 \langle \Psi_1 | T + U | \Psi_1 \rangle ,
\label{2}
\end{eqnarray}
where $H_1$  is the total Hamiltonian of the system in the
external field  $\varphi^{ext}_1 ({\bf r})$,   $T$  and  $U$ are
the operators of the kinetic energy and the energy of the
interparticle interaction of electrons, respectively. In this
case,  $\varphi^{ext}_2 ({\bf r}) \ne \varphi^{ext}_1 ({\bf r}) +
const$ is the potential of another external field to which the
ground state with wave function $\Psi_2$  and energy $E_2$
corresponds, which leads to the same density $n({\bf r)}$. The use
of the inequality similar to the strict inequality (1), but with
permutated indices, followed by summation of these inequalities,
results in a contradiction. This contradiction provides the lemma
proof to the contrary [1,9]. However, the strict inequality is a
direct consequence of the Rayleigh-Ritz minimum principle which,
as applied to quantum mechanics, is written as (see, e.g.,
[11,12])

\begin{eqnarray}
E_0\leq  \langle \Psi | H | \Psi \rangle , \qquad   \langle \Psi | \Psi \rangle =1,
\label{3}
\end{eqnarray}
where  $E_0$ is the ground state energy of the system of $N$  interacting electrons with Hamiltonian $H$   in an external field with
potential $\varphi^{ext}({\bf r)}$, and $\Psi $  is an arbitrary normalized wave function for the system of  $N$  electrons.
In this case, inequality (3) is not strict, which does not allow lemma proof.
The problem is what are the conditions of the transition from the non-strict inequality (3) to the strict inequality (1) which is the
basis for proving the Hohenberg-Kohn theorem. The first obvious condition is associated with the requirement of the ground state
non-degeneracy (see condition (b)). However, satisfying this condition is insufficient. It should be required that the ground state energy
$E_0$    would be a unique functional of the external field potential  $\varphi^{ext}({\bf r)}$,
\begin{eqnarray}
E_0 = E_0 \left( \left\{ \varphi^{ext} ({\bf r}) \right\} \right), \label{4}
\end{eqnarray}
i.e., as the external field potential $ \varphi^{ext}({\bf r})$
changes, the ground state energy  $E_0$ should take another value.
Currently, it seems impossible to prove this statement.

Probably, this problem is intimately related to the problem of
$\varphi$ - representability of the nonuniform density (see, e.g.,
[13,14]), which was also discussed by W. Kohn in its Nobel lecture
[9]\textbf{ and which has been solved in [15]. It should be
mentioned that if the inhomogeneous density $n({\bf r})$ does't
$\varphi^{ext} ({\bf r})$-representative the Hohenberg-Kohn
functional cannot be used (see [15] and the literature to this
paper). The problem considered in the present paper can be named,
by analogy with [15], the $\varphi^{ext} ({\bf
r})$-representability and $n({\bf r})$-representability of the
ground state energy. It is necessary to stress that the problem of
$\varphi^{ext} ({\bf r})$-representability and $n({\bf
r})$-representability of the ground state energy is primary for
proof of the Hohenberg-Kohn theorem.}

Thus, we come to the conclusion that the Hohenberg-Kohn lemma and
theorem are direct consequences of the statement that the ground
state energy of a system of interacting particles in an external
field is a unique functional of the potential of this field,
\begin{eqnarray}
E_0 = E_0 \left( N \left\{ \varphi^{ext} ({\bf r}) \right\} \right) \to E_0=E_0\left(  \left\{ n({\bf r}) \right\} \right), \qquad  N=
\int n({\bf r)}d{\bf r}.
\label{5}
\end{eqnarray}

The impossibility of proving the statement (4) does not cast doubt on the validity of the Hohenberg-Kohn lemma and theorem,
but makes it possible to obtain additional consequences under the assumption of their validity. In particular,
it follows from (4), (5), and the Hohenberg-Kohn lemma and theorem, that the nonuniform density $n({\bf r})$
corresponding to the ground state and the external field potential   $\varphi^{ext}({\bf r})$  are biunique functionals,
\begin{eqnarray}
\varphi^{ext}({\bf r}) = \varphi^{ext}\left(\left\{ n ({\bf r}) \right\} \right) \leftrightarrow
n ({\bf r}) =  n \left(\left\{ \varphi^{ext}({\bf r}) \right\} \right).
 \label{6}
\end{eqnarray}

We note that the Hohenberg-Kohn lemma and theorem can be proved
within quantum statistics for both the grand canonical ensemble by
analyzing corresponding inequalities for the thermodynamic
potential [16] and for the canonical ensemble by analyzing
corresponding inequalities for the free energy [17]. Thus, we can
repeat the above analysis, e.g., for the free energy $F$  of a
system of $N$ interacting electrons with Hamiltonian $H=H_0 + H_1$
at temperature $T$. In this case, the inequality (3) takes the
following form [18,19] (see also [20]):
\begin{eqnarray}
F \leq F_0 + \langle H_1 \rangle _0, \qquad F = T \ln Sp \left[ \exp \left( - \frac{H}{T}\right) \right],
\label{7}
\end{eqnarray}

\begin{eqnarray}
F_0 = -T \ln Sp \left[ \exp \left( - \frac{H_0}{T}\right) \right],  \qquad \langle H_1 \rangle_0 \equiv
Sp \left[ H_1\exp \left( - \frac{F_0-H_0}{T}\right) \right].
\label{8}
\end{eqnarray}

In relation (8), the Hamiltonian $H_1$ smallness is not assumed in any sense. Thus, to prove the Hohenberg-Kohn
lemma and theorem for quantum statistics, we come to the necessity of the requirement that the free energy $F$
of the system of $N$ interacting electrons at a given temperature $T$ was a unique functional of the external field
potential  $\varphi^{ext}({\bf r}) $  (see (4)),
\begin{eqnarray}
F=F\left(N,T,\left\{ \varphi^{ext} ({\bf r}) \right\} \right).
 \label{9}
\end{eqnarray}

In the limit  $T \to 0$, relation (9) obviously transforms into
relation (4). A statement similar to (9) can also be formulated
for the thermodynamic potential within the grand canonical
ensemble. Hence, when condition (9) is satisfied, statement (6)
that the nonuniform density  $n({\bf r}) $  and the external field
potential  $\varphi^{ext}({\bf r})$ are biunique functionals can
be generalized for equilibrium systems at a given temperature  $T$
within quantum statistics. Statement (6) makes it possible to
obtain relations important for the density functional, whose
importance is caused by the absence of the regular procedure of
the functional $F=F\left(T,\left\{ n ({\bf r}) \right\} \right)$
construction, in contrast to the case of the functional
$F=F\left(N,T,\left\{ \varphi^{ext} ({\bf r}) \right\} \right)$
which can be constructed using the perturbation theory diagram
technique (see, e.g., [21]), proceeding from the equivalence of
the grand canonical and canonical ensembles [22]. In particular,
it immediately follows from (6) that (see, e.g., [22])
\begin{eqnarray}
\int \frac {\delta n({\bf r}_1) }{\delta \varphi_{ext} ({\bf r}) } \frac{\delta \varphi_{ext} ({\bf r}) }{\delta n({\bf r}_2) } d{\bf r} =
\delta({\bf r}_1-{\bf r}_2) , \label{10}.
\end{eqnarray}

In turn, from the definitions of the free energy $F$  and
nonuniform density  $n({\bf r})$  (see (7), (8)), we find (see,
e.g., [21,22]) \textbf{the exact relations which are valid for the
arbitrary strong external potential $\varphi_{ext} ({\bf r})$}
\begin{eqnarray}
\frac {\delta F}{\delta \varphi_{ext} ({\bf r}) }= n({\bf r}) , \qquad \frac{\delta n ({\bf r}_1) }{\delta \varphi_{ext} ({\bf r}_2) }=
\chi({\bf r}_1,{\bf r}_2) , \label{11}
\end{eqnarray}
where $\chi({\bf r}_1,{\bf r}_2) $  is the so-called response function,
\begin{eqnarray}
\chi({\bf r}_1,{\bf r}_2) =\int_{0}^{1/T} \langle \Delta N ({\bf r}_1,\tau) \Delta N({\bf r}_2,0) \rangle d\tau, \qquad  A(\tau)=
\exp (H\tau)A \exp(-H\tau).
 \label{12}
\end{eqnarray}

Here  $\Delta N({\bf r})=N({\bf r})-n({\bf r})$, where  $N({\bf r})$  is the operator of the density of the number of particles.
The response function $\chi({\bf r}_1,{\bf r}_2) $  in the Coulomb system is directly related to the static permittivity
(see, e.g., [23]); in the classical system, it is related to the pair distribution function (see, e.g., [22]).
Furthermore, the response function can be calculated using the quantum field theory methods [21,23]. From (10) and (11), it immediately follows that
\begin{eqnarray}
\frac{\delta \varphi_{ext} ({\bf r}_1) }{\delta n({\bf r}_2)
}=\chi^{-1}({\bf r}_1,{\bf r}_2) \qquad \int \chi({\bf r}_1,{\bf
r})\chi^{-1}({\bf r},{\bf r}_2)d {\bf r}= \delta ({\bf r}_1-{\bf
r}_2).
 \label{13}
\end{eqnarray}
The second equation in (13) is used to calculate the inverse
response function $\chi^{-1}({\bf r}_1,{\bf r}_2)$  which is
directly related to the direct correlation function in the
classical system (see, e.g., [22]). Under the assumption that the
functional for the ground state energy  $E_0 \left(\left\{ n ({\bf
r}) \right\} \right)$ (or for the free energy $F \left(T,\left\{ n
({\bf r}) \right\} \right)$) is known, the definition of the
nonuniform density $n({\bf r})$  in the density functional theory
[1-9] is based on the minimum principle
\begin{eqnarray}
\delta E_0\left(\left\{ n ({\bf r}) \right\} \right)=0 \quad  ( \mbox{ or }  \delta F \left(T,\left\{ n ({\bf r}) \right\} \right)-0) \qquad
N= \int n({\bf r})d{\bf r}.
 \label{14}
\end{eqnarray}

The minimality condition (14) at a given number of particles  $N$, using the Legendre transform, can be written as
\begin{eqnarray}
\frac{\delta E_0\left(\left\{ n ({\bf r}) \right\} \right)}{\delta n({\bf r})}=
const \qquad \left(\mbox{ or } \frac{\delta F\left(T,\left\{ n ({\bf r}) \right\} \right)}{\delta n({\bf r})}= const \right )
.\label{15}
\end{eqnarray}

Taking into account the statement (6), it immediately follows from (15) that
\begin{eqnarray}
\frac{\delta F\left(\left\{ n ({\bf r}) \right\} \right)}{\delta n({\bf r})}=
 \int\frac{\delta F\left(T,\left\{ n ({\bf r}) \right\} \right)}{\delta \varphi_{ext} ({\bf r}_1)}
 \frac{\delta \varphi_{ext} ({\bf r}_1)} {\delta n({\bf r})}d {\bf r}_1  = const.
\label{16}
\end{eqnarray}

Using now (10) - (13), we find the condition which should be satisfied by the nonuniform density
$n ({\bf r})$ of the equilibrium system in the static external field,
\begin{eqnarray}
\int n({\bf r}_1)\chi^{-1}({\bf r}_1,{\bf r})d{\bf r}_1=const.
\label{17}
\end{eqnarray}

According to the above consideration, condition (17) can be
considered as a necessary condition of the Hohenberg-Kohn lemma
and theorem to be valid. In this case, the nonuniform density
$n({\bf r})$  is intimately related to the inverse response
function  $ \chi^{-1}({\bf r}_1,{\bf r}_2)$, so that the integral
in (17) is independent of the coordinate ${\bf r}$. Thus, the
statement (6) about the biunique correspondence between the
nonuniform density $n({\bf r})$ and the external field potential $
\varphi^{ext}({\bf r})$  in the equilibrium system is a necessary
condition for the validity of the Hohenberg-Kohn lemma and
theorem, whose direct consequence is the relation (17).
Furthermore, the statement (6) in principle makes it possible to
solve one of the main problems of the density functional theory,
associated with the absence of a regular procedure for determining
the functional $F=F \left(T,\left\{ n ({\bf r}) \right\} \right)$
itself for the free energy, or $E_0= E_0 \left(\left\{ n ({\bf r})
\right\} \right)$  for the ground state energy. Such a procedure
is based on the primary definition of an explicit form of the
functionals $F=F \left(N,T\left\{ \varphi^{ext} ({\bf r}) \right\}
\right)$ (or $ E_0= E_0 \left(N,\left\{ \varphi^{ext} ({\bf r})
\right\} \right)$) and $n(\{\varphi({\bf r})\})$, based on the
quantum field theory methods or other regular calculation methods.
On this basis, the functional  $ \varphi^{ext} ({\bf r})=
\varphi^{ext} \left(\left\{ n ({\bf r}) \right\} \right)$ is
determined, which is substituted into the found expressions for
the free energy   $F=F\left(N,T\left\{ \varphi^{ext} ({\bf r})
\right\} \right)$  (or for the ground state energy $ E_0= E_0
\left(N,\left\{ \varphi^{ext} ({\bf r}) \right\} \right)$ ). This
procedure is completely identical to the well-known procedure for
constructing the virial equation of state, based on the Mayer
diagram technique (see, e.g., [24]).

The authors (V.B. and S.T.) are thankful to the Netherlands Organization for
Scientific Research (NWO) for support of this work in the
framework of the grants on statistical physics and the Russian Foundation for Basic
Research for support in the framework of the projects no. 07-02-01464-a and no. 10-02-90418-Ukr-a.

\end{document}